\documentclass[doublecol]{epl2} 

\usepackage{graphicx}
\usepackage{amsmath}

\title{Two-component electron fluid in underdoped high-$T_c$ cuprate superconductors}
\shorttitle{Two-component electron fluid in underdoped cuprate superconductors} 

\author{J. G. Storey\inst{1} \and J. L. Tallon\inst{2}}
\shortauthor{Storey and Tallon}

\institute{                    
  \inst{1} MacDiarmid Institute for Advanced Materials and Nanotechnology - School of Chemical and Physical Sciences, Victoria University of Wellington, P.O. Box 500, Wellington, New Zealand.\\
  \inst{2} MacDiarmid Institute - Industrial Research Ltd., P.O. Box 31310, Lower Hutt, New Zealand.
}

\date{\today}

\abstract{
Evidence from NMR of a two-component spin system in cuprate high-$T_c$ superconductors is shown to be paralleled by similar evidence from the electronic entropy so that a two-component quasiparticle fluid is implicated. We propose that this two-component scenario is restricted to the optimal and underdoped regimes and arises from the upper and lower branches of the reconstructed energy-momentum dispersion proposed by Yang, Rice and Zhang (YRZ) to describe the pseudogap. We calculate the spin susceptibility within the YRZ formalism and show that the doping and temperature dependence reproduces the experimental data for the cuprates.
}

\begin{document}

\maketitle


From the electronic entropy we present evidence for a two-component electron fluid in underdoped high-$T_c$ cuprates matching similar evidence from NMR \cite{HAASE1,HAASE2,MEISSNER}. We then show that this two-component behavior probably arises from Fermi surface reconstruction and the associated band-splitting that occurs due to pseudogap correlations. We illustrate this using the model proposed by Yang, Rice and Zhang (YRZ) \cite{YRZ}. Implicit in this interpretation is the prediction that single-component behavior is recovered in the overdoped region where the pseudogap closes around $p\approx0.19$ holes/Cu.

Hole-doped HTS, in their overdoped state, possess a large Fermi surface \cite{HUSSEY,PLATE,VIGNOLLE} enclosing $(1+p)$ carriers where $p$ is the doped hole concentration residing largely on in-plane oxygen orbitals and the ``1" arises from the unpaired electrons residing on the Cu sites. This naively suggests a two-component electron system. However, Zhang and Rice \cite{ZRS} showed that the doped holes form a local singlet state with the hole on the Cu 3$d_{x^2-y^2}$ orbital, the so-called Zhang-Rice singlet. It was suggested that the singlet binding energy was so large that the oxygen 2$p$ holes do not contribute to the spin susceptibility thus resulting in a single-component spin scenario. 
This situation has been considered long-established starting from Takigawa {\it et al}. \cite{TAKIGAWA}, who showed that the $^{17}$O and $^{63}$Cu Knight shifts, $^{17}K(T)$ and $^{63}K(T)$ respectively, for underdoped YBa$_2$Cu$_3$O$_{7-\delta}$ exhibit an identical temperature dependence. Subsequent $^{89}$Y NMR studies confirmed the same $T$-dependence in $^{89}K(T)$ \cite{ALLOUL} adding further weight to the single-component scenario.

\begin{figure*}
\centering
\includegraphics[width=59mm]{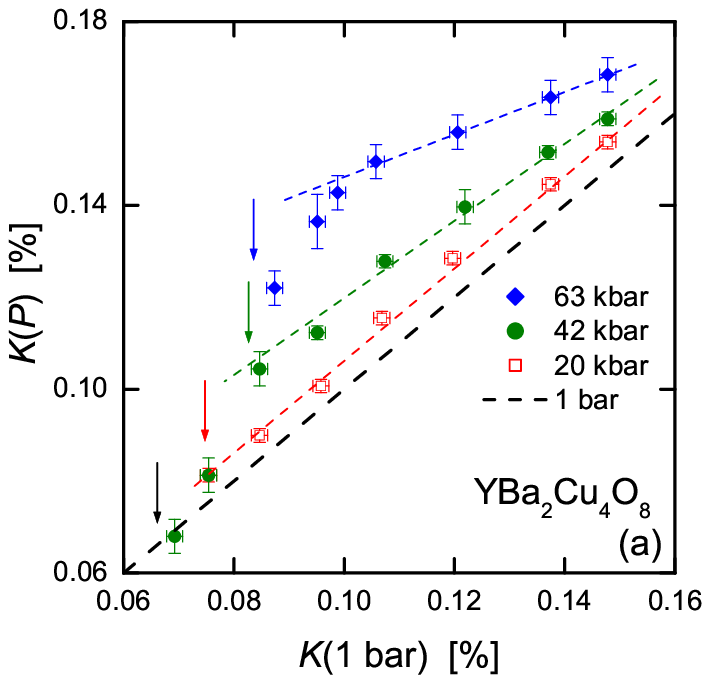}%
\hspace{1mm}%
\includegraphics[width=56mm]{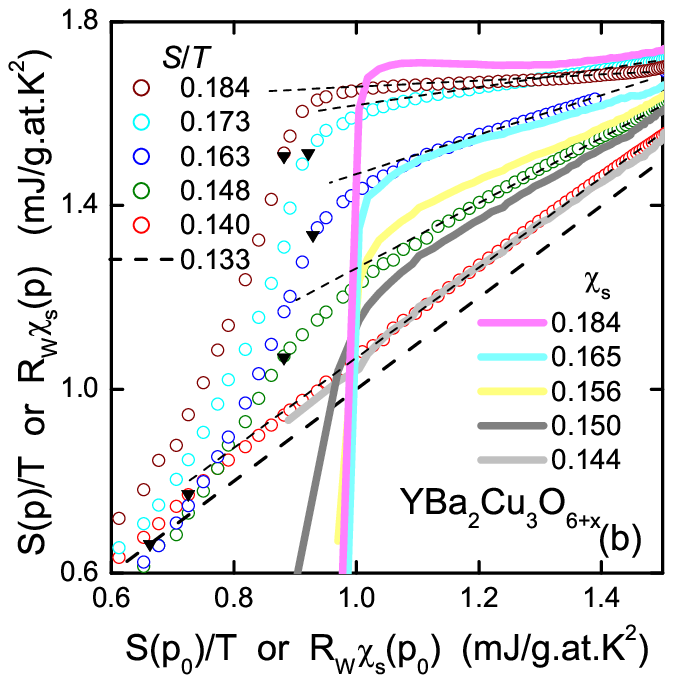}
\hspace{1mm}%
\includegraphics[width=58mm]{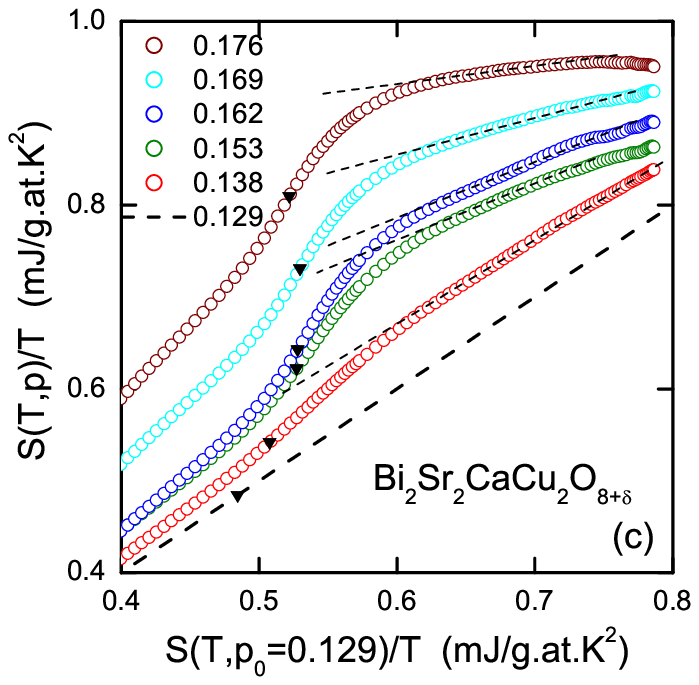}
\caption{\small
(Color online) (a) Reproduced from ref. [9]: the Knight shift, $K(P)$, for YBa$_2$Cu$_4$O$_8$ at $P=20$, 42 and 63 kbar plotted vs $K(1$ bar) where $T$ is the implicit variable. Arrows indicate $T_c$. Above the range of SC fluctuations $K(P)$ is linear in $K(1$ bar). 
(b) An analogous plot of $S(T,p)/T$ for YBa$_2$Cu$_3$O$_{6+x}$ plotted vs $S(T,p_0=0.133)/T$ for $x=$ 0.76, 0.80, 0.87, 0.92 and 0.97 (open circles). The corresponding $p$ values are listed. Also shown is $R_W \chi_s(T,p)$ vs $R_W \chi_s(T,p_0=0.135)$, expressed in entropy units (solid curves). (c) $S(T,p)/T$ for Bi$_2$Sr$_2$CaCu$_2$O$_{8+\delta}$ plotted vs $S(T,p_0=0.129)/T$ for the listed $p$-values. 
} \label{KPvsK0}
\end{figure*}

However, this \revision{single-component} picture was recently questioned by Haase {\it et al} \cite{HAASE1,HAASE2,MEISSNER}. In order to eliminate the Meissner term in the spin shift they constructed a term $G_\bot$ by subtracting the Knight shift for the apical oxygen with field perpendicular to the $c$-axis from that for the planar Cu shift, and similarly $G_{\parallel}$ for field parallel to the $c$-axis. Thus,
\begin{equation}
\begin{array}{ll}
G_{\bot}(T) & = \,^{63,\bot}\!K(T) - ^{17,A,\bot}\!K(T), \\
\\
G_\parallel(T) & = \,^{17,P,\parallel}\!K(T) - ^{17,A,\parallel}\!K(T),
\end{array}
\label{Gparl}
\end{equation}
\noindent where the superscript A refers to the apical oxygen while the superscript P refers to planar oxygen.

These authors showed that $G_\bot(T)$ displayed a $T$-independent Pauli-like metallic susceptibility, while $G_{\parallel(T)}$ displayed a strongly $T$-dependent susceptibility consistent with a gapped normal-state spectrum. By constructing a two-component ansatz:
\begin{equation}
\begin{array}{ll}
G_{\bot}(T) & = c_{11}\chi_1 + c_{12}\chi_2,\\
\\
G_\parallel(T) & = c_{21}\chi_1 + c_{22}\chi_2,\\
\end{array}
\label{ansatz}
\end{equation}
\noindent where $\chi_1$ and $\chi_2$ are the two putative uniform susceptibilities, Haase {\it et al.} \cite{HAASE1} showed that $\chi_1(T)$ has a characteristic pseudogap-like $T$-dependence, falling steadily towards zero with reducing $T$, while $\chi_2(T)$ has a constant Pauli-like behavior and only falls to zero below $T_c$ with the opening of the superconducting (SC) energy gap.

In introducing $\chi_1$ and $\chi_2$ Haase {\it et al.} took a lead from Johnston \cite{Johnston} who, from static bulk measurements, inferred a two-component susceptibility of the form $\chi(p,T) = \chi_1(p,T) + \chi_2(p)$ where $\chi_2$ is a function of doping only. However, the features he sought to explain arise naturally from the presence of the van Hove singularity (vHs) in the overdoped region which ensures an overall increasing density of states across most of the phase diagram (as reflected in the $\chi_2(p)$ term). A single-component susceptibility fully accounts for the complete evolution of $^{89}K_s$ across the underdoped and overdoped regions in Y$_{0.8}$Ca$_{0.2}$Ba$_2$Cu$_3$O$_{7-\delta}$ \cite{STOREYPHASE} provided the full ARPES derived dispersion is utilized, including the vHs. It requires a differential measurement, as in Eq.(1), to expose the true underlying two-component behavior.

To make the problem more specific we consider the most recent study from the Haase group \cite{MEISSNER}, which also supports the two-component picture. In this they carried out high-pressure diamond-anvil NMR measurements on YBa$_2$Cu$_4$O$_8$ to 63 kbar. Pressure, $P$, acts dominantly, though not exclusively, to increase the doping by transferring holes from the Cu$_2$O$_2$ chains to the CuO$_2$ planes. Thus one typically observes a pressure-induced decrease in thermoelectric power \cite{ZHOUJS} consistent with an increase in hole concentration, $p$, resulting in a pressure-induced increase in $T_c$ on the underdoped side and a decrease in $T_c$ on the overdoped side \cite{SCHLACHTER}. (That the effect of pressure is not {\it merely} to increase the hole concentration is evident from the fact that the maximum $T_c$ is increased from approx 93 K at ambient pressure to approximately 107 K at a pressure of about 7 GPa \cite{SCHOLTZ}). Consistent with this, Meissner {\it et al.} \cite{MEISSNER} observed $^{17}K(T)$ to progress from a strongly $T$-dependent behavior under ambient pressure, typical of a pseudogapped underdoped cuprate, towards a nearly $T$-independent behavior at a pressure of 63 kbar, more typical of an optimally-doped cuprate. By plotting $^{17}K(T,P)$ versus $^{17}K(T,P=$ 1 bar$)$ for $P=20$, 42 and 63 kbar with $T$ as the implicit variable they obtained a linear relation that showed a characteristic progression with increasing $P$. This plot is reproduced in Fig.~\ref{KPvsK0}(a).

The authors draw attention to the fact that well above $T_c$ there is a linear region that for the lowest doping extends down almost to $T_c$. They model this linear behavior within a two-component scenario.

Firstly we wish to point out that identical behavior is seen in the electronic entropy, $S(T)$, thus indicating that it is not just the spin system that exhibits two components but the total quasiparticle ensemble.

For a nearly free-electron system the spin susceptibility, $\chi_s$ and $S/T$ are related via the Wilson ratio, $R_W$:
\begin{equation}
R_W \chi_s(T) = S(T)/T.
\label{Wilson}
\end{equation}

Accordingly, the open circles in Fig.~\ref{KPvsK0}(b) and (c) show $S(T,p)/T$ versus $S(T,p_0)/T$ (where $T$ is the implicit variable) for YBa$_2$Cu$_3$O$_{6+x}$ (b) and for Bi$_2$Sr$_2$CaCu$_2$O$_{8+\delta}$ (c). We choose $p_0 = 0.133$ for the former and $p_0 = 0.129$ for the latter, which are close to the zero-pressure doping state of YBa$_2$Cu$_4$O$_8$, $p_0\approx0.13$. The same generic behavior also occurs in plots of $S(T,p)/T$ versus $S(T,p_0)/T$ for Y$_{0.8}$Ca$_{0.2}$Ba$_2$Cu$_3$O$_{6+x}$ (not shown). The correspondence between $K(P)$ and $S(p)/T$ is remarkable.

Assuming the major effect of pressure is to increase doping and letting $\alpha$ be the thermopower, then $(\partial\alpha/\partial P)_{T=290} = -0.075 \mu$V K$^{-1}$kbar$^{-1}$ \cite{ZHOUJS}. Using the thermopower correlation with doping $(\partial\alpha/\partial p)_{T=290} = -134 \mu$V K$^{-1}$hole$^{-1}$ \cite{COOPER} one finds $(\partial p/\partial P) = 5.6 \times 10^{-4}$ holes/kbar. Thus the doping states of YBa$_2$Cu$_4$O$_8$ at 20, 42 and 63 kbar are 0.141, 0.154 and 0.165 holes/Cu. Visually the red, green and blue data sets in Fig.~\ref{KPvsK0}(a) correspond closely to the red, green and blue data sets in Fig.~\ref{KPvsK0}(b) and so should be at roughly the same doping states. This is indeed the case where in Fig.~\ref{KPvsK0}(b) the dopings are seen to be 0.140, 0.148 and 0.163 holes/Cu.

Moreover, we show that the correspondence between $\chi_s$ and $S/T$ is in excellent quantitative agreement with Eq.~\ref{Wilson}. The solid curves in Fig.~\ref{KPvsK0}(b) show values of $\chi_s(T,p)$ versus $\chi_s(T,p_0=0.135)$ expressed in entropy units using the Wilson ratio for nearly-free electrons $R_W^0 = (\pi^2/3)(k_B/\mu_B)^2$. We use the bulk susceptibility $\chi_s$ data of Cooper and Loram \cite{COOPER}. One can see that the correspondence between $\chi_s$ and $S/T$ is not just qualitative but quantitative. These plots reveal precisely the same magnitudes as $S/T$ and the same breakaway from linear behavior sets in well above $T_c$ due to strong SC fluctuations \cite{TALLONFLUCS}. (Note that below $T_c$ the diamagnetism leads to different behavior from $S/T$). We thus conclude that there is in fact a two-component {\it quasiparticle system}, not just a two-component {\it spin system}.

We now turn to our central thesis that this generic behavior, seen in both the spin susceptibility and the entropy, arises from band splitting that occurs when the Fermi surface reconstructs due to competing orders such as a CDW, SDW or short-range AF correlations. We illustrate this within the YRZ scenario where the electron self-energy term $E_g^2(\textbf{k})/(\omega+\xi_\textbf{k}^0)$ reconstructs the $E(\textbf{k})$ dispersion into upper and lower branches which yield our two-component quasiparticle ensemble. Here $\xi_\textbf{k}^0 = -2t(p)(\cos k_x + \cos k_y)$ is the nearest-neighbor term in the tight-binding $E(\textbf{k})$ dispersion. The coherent part of the electron Green's function is given by:
\begin{equation}
G(\textbf{k},\omega,p) = g_t(p)\left[\omega - \xi_\textbf{k} - \frac{E_g^2(\textbf{k})}{\omega+\xi_\textbf{k}^0}\right]^{-1}.
\label{Greens}
\end{equation}

\noindent where $\xi_k = -2t(p)(\cos k_x + \cos k_y) - 4t'(p)\cos k_x \cos k_y - 2t''(p)(\cos2k_x + \cos2k_y) - \mu_p(p)$ is the tight-binding dispersion to third-nearest neighbors, and $E_g(\textbf{k}) = \frac{1}{2}E_g^0(p)(\cos k_x - \cos k_y)$ is the pseudogap with doping dependence $E_g^0(p) = 3t_0(0.2 - p)$, while for $p > 0.2$ we have $E_g^0(p) = 0$. (This means the pseudogap closes at $p=0.2$ however we have extensively shown this to occur at slightly lower doping $p=0.19$\cite{OURWORK1}. Here we retain the value 0.2 to remain consistent with YRZ). The chemical potential $\mu_p(p)$ is chosen according to the Luttinger sum rule. The doping dependent coefficients are given by $t(p)=g_t(p)t_0+(3/8)g_s(p)J\chi$, $t^\prime(p)=g_t(p)t_0^\prime$ and $t^{\prime\prime}(p)=g_t(p)t_0^{\prime\prime}$, where $g_t(p)=2p/(1+p)$ and $g_s(p)=4/(1+p)^2$ are the Gutzwiller factors. The bare parameters $t^\prime/t_0=-0.3$, $t^{\prime\prime}/t_0=0.2$, $J/t_0=1/3$ and $\chi=0.338$ are the same as used previously\cite{YRZ}.

Equation~\ref{Greens} can be re-written as
\begin{equation}
G(\textbf{k},\omega,p)=\sum_{\alpha=\pm}{\frac{g_t(p)W_\textbf{k}^\alpha(p)}{\omega-E_\textbf{k}^\alpha(p)}} ,
\label{eq:GYRZ2}
\end{equation}
where the energy-momentum dispersion is reconstructed by the pseudogap into upper and lower branches
\begin{equation}
E_\textbf{k}^\pm=\frac{1}{2}(\xi_\textbf{k}-\xi_\textbf{k}^0)\pm\sqrt{\left(\frac{\xi_\textbf{k}+\xi_\textbf{k}^0}{2}\right)^2+E_g^2(\textbf{k})} ,
\label{eq:EK}
\end{equation}
which are weighted by
\begin{equation}
W_\textbf{k}^\pm=\frac{1}{2}\left[1\pm\frac{(\xi_\textbf{k}+\xi_\textbf{k}^0)/2}{\sqrt{[(\xi_\textbf{k}+\xi_\textbf{k}^0)/2]^2+E_g^2(\textbf{k})}}\right] .
\label{eq:WK}
\end{equation}
The spectral function is given by
\begin{equation}
A(\textbf{k},\omega,p)=\sum_{\alpha=\pm}{g_t(p)W_\textbf{k}^\alpha\delta(\omega-E_\textbf{k}^\alpha)} ,
\label{eq:Akw}
\end{equation}
from which the density of states can be calculated
\begin{equation}
N(\omega)=\sum_{\textbf{k}}{A(\textbf{k},\omega)} .
\label{eq:DOS}
\end{equation}
Finally, the spin susceptibility is given by
\begin{equation}
\chi_s = 2\mu_B^2 \int {\left(-\frac{\partial f}{\partial \omega}\right) N(\omega)} d\omega.
\label{eq:CHI}
\end{equation}

\begin{figure*}
\centering
\includegraphics[trim=0mm 2mm 0mm 0mm, width=58mm]{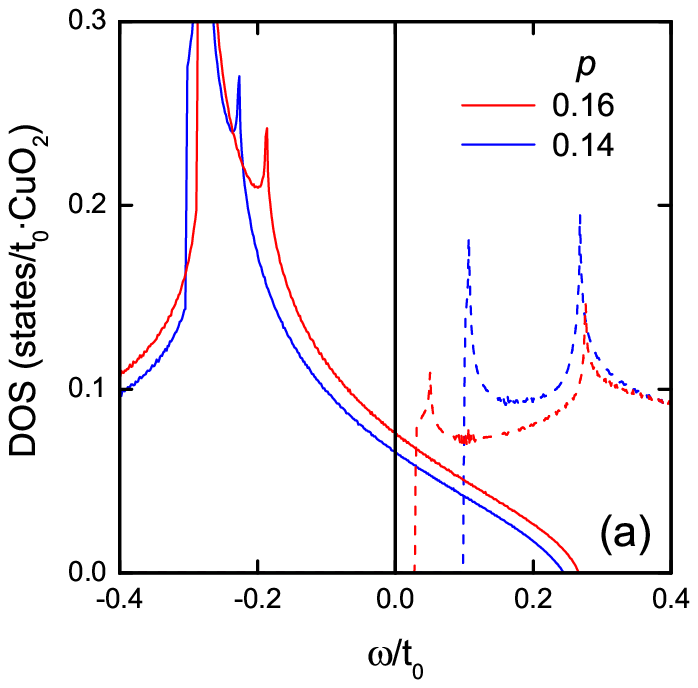}%
\hspace{1mm}%
\includegraphics[trim=0mm 0mm 0mm 0mm, width=58mm]{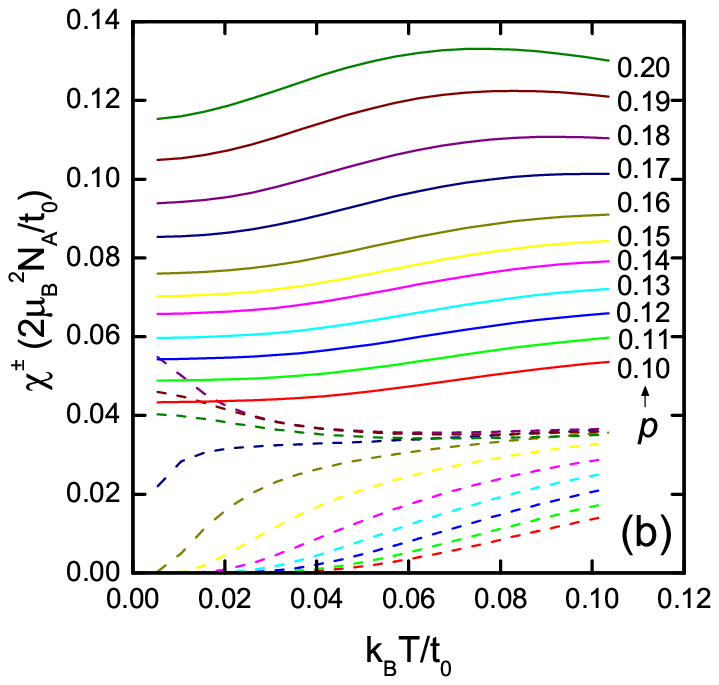}
\hspace{1mm}%
\includegraphics[trim=0mm 1mm 0mm 0mm, width=58mm]{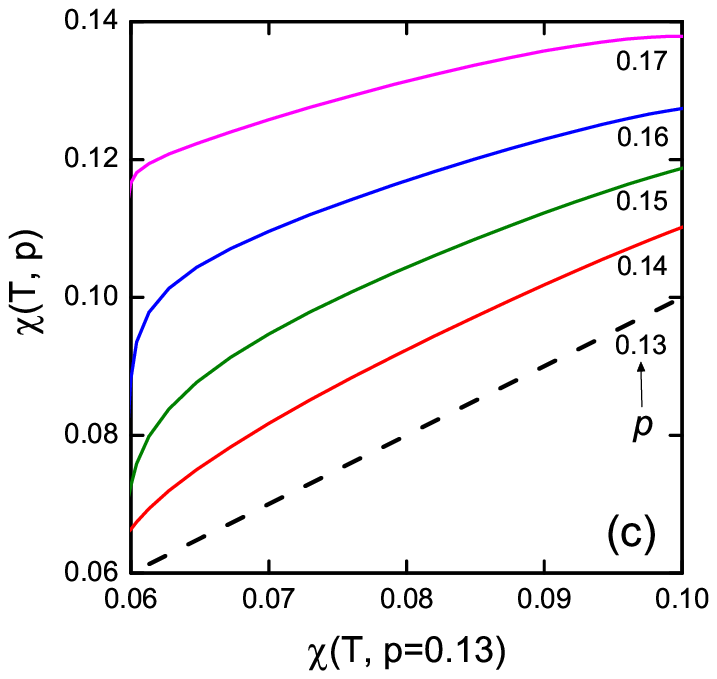}
\caption{\small
(Color online) (a) Partial density of states calculated from the lower (solid) and upper (dashed) branches of the YRZ reconstructed dispersion. (b) Susceptibility calculated from the lower (solid) and upper (dashed) partial density of states. (c) Total $\chi_s(T,p)$ plotted versus $\chi_s(T,p_0=0.13)$ where $T$ is the implicit variable (units of $\chi$ are the same as in (b)). } \label{FIG2}
\end{figure*}

Figure~\ref{FIG2}(a) shows the partial densities of states (PDOS) for $p=0.14$ and 0.16 calculated from the upper and lower branches of the reconstructed dispersion given by Eq.~\ref{eq:EK}. The PDOS of the lower branch is quasi-linear across the Fermi level ($\omega=0$) resulting in a roughly $T$-independent contribution to the susceptibility at low doping, shown in Fig.~\ref{FIG2}(b) by the solid curves. For dopings below about 0.18, the PDOS of the upper branch lies above the Fermi level resulting in a gapped spectrum. With decreasing doping the upper PDOS is pushed further from $\omega=0$, producing a $T$-dependent contribution to $\chi$ that is characteristic of the pseudogap state, shown in Fig.~\ref{FIG2}(b) by the dashed curves.

The two sets of susceptibilities shown by the dashed and solid curves in Fig.~\ref{FIG2}(b) are to be directly compared with the functions $\chi_1(T)$ and $\chi_2(T)$, respectively, reported by Haase {\it et al.,} \cite{HAASE1,HAASE2}. They reveal the same qualitative behavior but there are points of difference, primarily in their relative magnitudes where, at high temperature, the lower-branch susceptibility is more than three times the magnitude of the upper-branch susceptibility. By contrast, Haase {\it et al.} find $\chi_2$ exceeds $\chi_1$ by around 50\% at 300 K. We return to this discrepancy below.

In order to compare directly with the pressure- and doping-dependent data shown in Figure~\ref{KPvsK0} we plot in Fig.~\ref{FIG2}(c) the sum of the susceptibilities (from the upper and lower branches) as a function of the susceptibility sum for $p=0.13$, with $T$ as the implicit variable. The qualitative features seen in the normal-state Knight shift (Fig.~\ref{KPvsK0}(a)) and electronic entropy (Figs.~\ref{KPvsK0}(b) \& (c)) are reproduced in detail. In particular, the high-$T$ downturn seen for $p=0.17$ reflects the proximity of the vHs. The same downturn is seen in the Bi$_2$Sr$_2$CaCu$_2$O$_{8+\delta}$ data (Fig.~\ref{KPvsK0}(c)) where the vHs indeed lies nearby \cite{2212VHS} in the moderately overdoped region, whereas it is not evident in the data for YBa$_2$Cu$_3$O$_{7-\delta}$ (Fig.~\ref{KPvsK0}(b)) where the vHs is known to lie in the more deeply overdoped region.

We are however left with two remaining questions: (i) why does $G_{\bot}(T)$ reflect more the lower branch susceptibility while $G_{\parallel}(T)$ reflects more the upper branch susceptibility? And (ii) there is the question, mentioned above, of the relative magnitudes of the two susceptibilities. We suggest these have a common origin, as follows:

The apical oxygen is coupled to the planar copper and oxygen orbitals via the Cu 4$s$ orbital \cite{Xiang}. This introduces matrix elements that weight the $\textbf{k}$-space sums, minimizing the contribution along the zone diagonals and maximizing contributions at the ($\pi$,0) zone boundaries. This clearly will diminish the contribution to the susceptibility from the lower branch, effectively reducing the magnitude of the coefficients $c_{12}$ and $c_{22}$. As a consequence the relative contributions of $\chi_1$ and $\chi_2$ to $G_{\bot}(T)$ and $G_{\parallel}(T)$ differ, with $G_{\bot}(T)$ dominated more by $\chi_2$. The $c$-axis hopping matrix is $t_\bot(\textbf{k}) = t_\bot^0 \mu_{\textbf{k}}^2$ where $\mu_{\textbf{k}} = 0.5\, \omega_{\textbf{k}}\, (\cos k_x - \cos k_y)$. We find this does indeed bring the magnitudes of $\chi_1$ and $\chi_2$ closer together for lower doping but not so much at higher doping close to where the pseudogap closes. We await a more rigorous treatment of the precise interaction of the apical nucleus with the two global spin susceptibilities.

In conclusion, we show that the entropy term $S(T)/T$ displays the same doping evolution as the pressure-dependent Knight shift, thus indicating that the two-component electronic behavior resides in the quasiparticle spectrum and not just in the spin spectrum. We then show that the essential features of the two-component system are likely to arise from band splitting due to zone-folding effects as described e.g. by the Yang-Rice-Zhang model for the pseudogap. The pseudogap-like susceptibility $\chi_1$ inferred by Haase {\it et al.} arises from the upper branch and the Pauli-like $\chi_2$ arises from the lower branch. If correct then it follows that single-component electronic behavior will be recovered when the pseudogap closes. Measurements of $G_{\bot}$ and $G_{\parallel}$ will then help settle the still contentious issue as to whether the pseudogap closes abruptly, both above the SC dome and below it at a putative quantum critical point at $p \approx 0.19$.

\acknowledgments

\end{document}